\documentclass[twocolumn,nofootinbib]{revtex4}

\usepackage{graphicx}
\usepackage{epsfig}

\begin{document}

\title{Dynamical system analysis of interacting models}

\author{S. Carneiro and H. A. Borges}

\affiliation{Instituto de F\'{\i}sica, Universidade Federal da Bahia, 40210-340, Salvador-BA, Brazil}

\begin{abstract}
We perform a dynamical system analysis of a cosmological model with linear dependence between the vacuum density and the Hubble parameter, with constant-rate creation of dark matter. We show that the de Sitter spacetime is an asymptotically stable critical point, future limit of any expanding solution. Our analysis also shows that the Minkowski spacetime is an unstable critical point, which eventually collapses to a singularity. In this way, such a prescription for the vacuum decay not only predicts the correct future de Sitter limit, but also forbids the existence of a stable Minkowski universe. We also study the effect of matter creation on the growth of structures and their peculiar velocities, showing that it is inside the current errors of redshift space distortions observations. 
\end{abstract}

\maketitle

\section{Introduction}

The access to a large quantity and variety of astronomical data has lead cosmology to the level of an observational based discipline, in which questions can be answered on a physical ground, and speculative models are more and more constrained. In fact, some old issues have already been physically treated since the advent of inflation. For example, the flatness and horizon problems, which could only be understood in terms of a fine-tuning of initial conditions, can now be posed as the consequence of a dynamical process. There are, however, some issues for which the answers are still rather speculative. The deeper one is, perhaps, why spacetime is what it is. Why is it expanding and accelerating, not an empty and eternal Minkowski spacetime? Is it approaching an asymptotic de Sitter regime, or is a different destiny allowed? What determines the energy scale of this de Sitter phase? If spacetime has been always expanding, is there any (standard) physical process that could avoid the initial singularity?

Surprisingly enough, possible answers for some of these issues may be found in the context of relatively well established physics, like quantum field theories in curved spacetime. As we will see, the QCD vacuum, although hardly fully understood, may play a relevant role. In a spacetime with physical horizon distance $1/H$, the vacuum density of a massless scalar field (or a massive field in the high energy limit, when its mass can be neglected) can be estimated as $\rho_V \sim E H^3$, with the fluctuation energy $E$ evaluated with the help of the uncertainty relation $E/H \sim 1$, which leads to $\rho_V \sim H^4$. This heuristic estimation is corroborated by conformal field theories results \cite{conformal}, and we will explore its consequences in Section \ref{highz}. However, in the case of a highly non-linear and strongly interacting theory like the low energy QCD, fluctuations are confined to a volume $1/m^3$, where $m$ is the energy scale of the chiral phase transition, and the vacuum density should be given instead by $\rho_V \sim m^3 H$, a result also suggested by QFT estimations \cite{Schutzhold}\footnote{There is a theorem according to which the vacuum density can only be proportional to even powers of $H$. However, it is valid for non-interacting fields and under the assumption that the vacuum energy-momentum tensor is conserved \cite{Wald}. None of these two conditions are fulfilled here.}. Supposing an approximately de Sitter spacetime, we have $\rho_V \sim H^2$, and hence $\rho_V \sim m^6$. Using $m \approx 150$ MeV, we obtain the order of magnitude of the observed cosmological constant.

If spacetime is not exactly de Sitter, a vacuum density linearly proportional to $H$ will vary. We will see from Eq. (\ref{conservation}) below that, if we adopt for vacuum the covariant equation of state $p_V = - \rho_V$, such a variation necessarily implies an interaction between vacuum and matter. Interacting models, linear or not, have been extensively studied in the literature for a long time, with different motivations \cite{iModels}. The linear model we are considering here was presented in Ref. \cite{Humberto} and has been tested against precise observations, showing a good concordance and alleviating some tensions that appear in the standard $\Lambda$CDM model \cite{linearmodel,hermano}. After the usual radiation and matter-dominated phases, it presents a late-time epoch dominated by a decaying vacuum, with concomitant creation of dark matter, which tends to de Sitter.

In the present paper we shall perform a dynamical system analysis of the linear model, from which some new aspects will be evidenced\footnote{For a dynamical system analysis of a more general class of interacting models, see \cite{arevalo}.}. First, we will show that the de Sitter regime is in fact an asymptotically stable critical point, future limit of any expanding spacetime. Secondly, and maybe more interesting, our analysis shows that the Minkowski spacetime is also a critical point, but unstable, which inevitably collapses to a singularity. In this way, the above prescription for the QCD vacuum condensate, if correct, not only predicts the correct future de Sitter limit, but also forbids the existence of a stable Minkowski universe.

\section{Dynamical system analysis}

\begin{figure*}
\centerline{\includegraphics[height=4.cm]{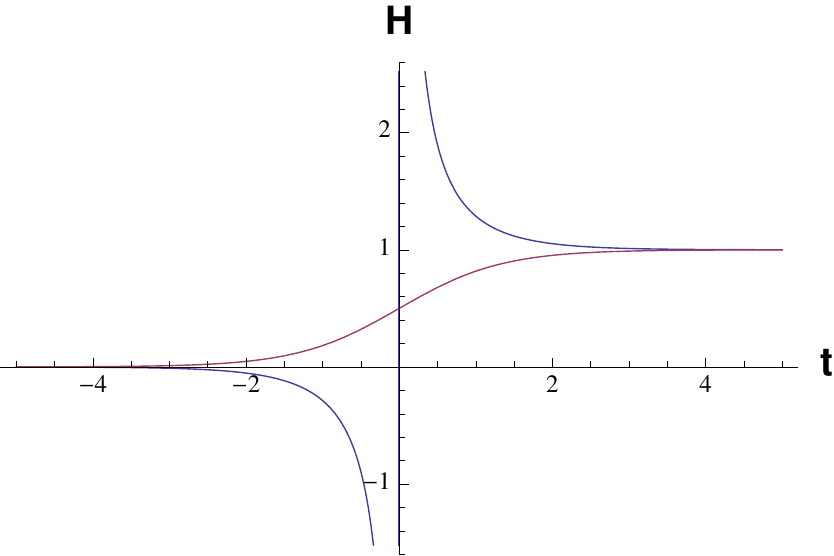} \hspace{.4in} \includegraphics[height=4.cm]{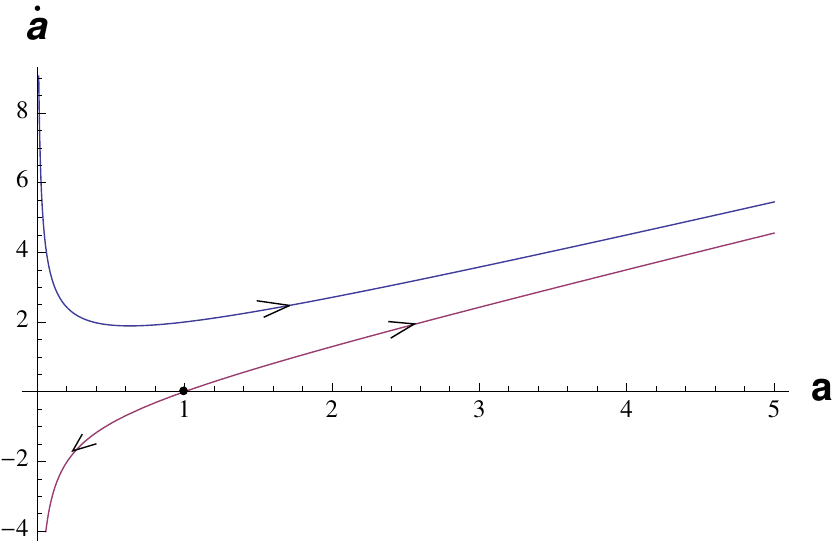}}
\caption{{\bf Left:} $H$ as a function of $t$ in the linear model. {\bf Right:} The phase space $\dot{a} \times a$.}
\end{figure*}

\label{iModel}

Let us consider a spatially flat FLRW spacetime fulfilled with pressureless dark matter of density $\rho_m$, and take for the vacuum component $\rho_V = \sigma H$ ($\sigma >0$ and constant) and $p_V = - \rho_V$. The Friedmann equations are
\begin{eqnarray} \label{Friedmann}
3H^2 &=& \rho_m + \rho_V,\\
\label{conservation}
\dot{\rho}_m + 3 H \rho_m &=& \Gamma \rho_m = - \dot{\rho}_V,
\end{eqnarray}
where $\Gamma$ defines the rate of matter creation. It is easy to show that $\Gamma = \sigma/2$ and, therefore, dark matter is created at a constant rate. From (\ref{Friedmann})-(\ref{conservation}) we can derive the Raychaudhuri equation
\begin{equation} \label{1st}
3H^2 - \sigma H + 2 \dot{H} = 0.
\end{equation}
The critical points are given by
\begin{equation}
\dot{H} = 0 \quad \rightarrow \quad H = \sigma/3 \quad \text{or} \quad H = 0.
\end{equation}
The first corresponds to a de Sitter spacetime, and the second, to the Minkowski spacetime.

We can rewrite equation (\ref{1st}) as $\dot{H} = g(H)$,
where
\begin{equation}
g(H)  = \frac{\sigma}{2} H - \frac{3}{2} H^2.
\end{equation}
Hence,
\begin{equation}
g'(H) = \frac{\sigma}{2} - 3H.
\end{equation}
At the de Sitter critical point, we have $g'(\sigma/3) = -\sigma/2 < 0$. This means that the de Sitter solution is asymptotically stable \cite{ODE} (any trajectory with initial condition $H > \sigma/3$ tends to it). On the other hand, for the Minkowski critical point we have $g'(0) = \sigma/2 > 0$. This means that this is an unstable solution (any trajectory with $H < 0$ collapses to a singularity, as shown below). 

In order to understand these results, let us find the general solution of (\ref{1st}). It is given by
\begin{equation} \label{solution}
H = \frac{\sigma/3}{1 \pm e^{-\sigma t/2}},
\end{equation}
where the origin of time was appropriately chosen.
The behavior of this solution is shown in the left panel of Fig. 1 (where we have fixed $\sigma = 3$ for convenience). We can see that, indeed, $H = \sigma/3$ is a stable critical point, while $H = 0$ is unstable. 
Note that the red curve on the left panel, which represents a spacetime that begins as Minkowski and tends to de Sitter, and corresponds to the positive-sign branch of (\ref{solution}), is physically meaningless. Indeed, from the Friedmann equations above we can show that $\rho_m = -2\dot{H}$ and, therefore, an increasing $H$ would correspond to a negative matter density, violating in this way the weak energy condition\footnote{On the other hand, we are assuming that, for a collapsing universe (the negative-$H$ branch on the left panel of Fig. 1), the vacuum density $\rho_V = \sigma H$ is negative. It must be, if we believe that the vacuum condensate decays into incoherent dark matter particles, and not the opposite (i.e. $\Gamma > 0$, see Eq. (\ref{conservation})).}.

Let us now define the phase-space variables $x_1 = a$ and $x_2 = \dot{a}$. We have, from (\ref{1st}),
\begin{eqnarray} \label{sistema1}
\dot{x}_1 &=& x_2,\\ \label{sistema2}
\dot{x}_2 &=& \frac{\sigma}{2} x_2 - \frac{1}{2} \frac{x_2^2}{x_1}.
\end{eqnarray}
The critical points $H = \sigma/3$ and $H = 0$ correspond, respectively, to $x_2/x_1 = \sigma/3$ and $x_2 = 0$.
For $H = 0$, the determinant of the Jacobian matrix $J_{ij} = \frac{\partial \dot{x}_i}{\partial x_j}$ is zero, that is, in the vicinity of this point the system above can be reduced to a single equation, with solution
\begin{equation}
(a,\dot{a}) = C_1 \vec{k}_1 + C_2 \vec{k}_2 e^{\sigma t/2},
\end{equation}
where $\vec{k}_1 = (1,0)$ and $\vec{k}_2 = (1,\sigma/2)$. The critical point (approached as $t \rightarrow -\infty$) is degenerated: $(a,\dot{a}) = (a,0)$, reflecting the fact that the scale factor $a$ can have any (positive) value in the Minkowski spacetime.

For $H = \sigma/3$, the eigenvalues of the Jacobian matrix, calculated at the critical point, are given by the characteristic equation \begin{equation}
\lambda^2 - \frac{\sigma}{6} \lambda - \frac{\sigma^2}{18} = 0.
\end{equation}
Its roots are $\lambda = \sigma/3$ and $\lambda = - \sigma/6$, i.e., the eigenvalues are real and have opposite signs. This means that the critical point is hyperbolic (a saddle point) \cite{ODE}. Any phase-space trajectory in its vicinity can be written as
\begin{equation}
(a,\dot{a}) = C_1 \vec{k}_1 e^{\sigma t/3} + C_2 \vec{k}_2 e^{-\sigma t/6},
\end{equation}
where $\vec{k}_1 = (1,\sigma/3)$ and $\vec{k}_2 = (1,-\sigma/6)$. For any $C_1 \neq 0$, it tends to the de Sitter point $H = \sigma/3$ for $t \rightarrow \infty$.

In the right panel of Fig. 1 we show typical oriented trajectories obtained by solving the system (\ref{sistema1})-(\ref{sistema2}), which general solution has the form
\begin{equation}
x_2 = x_1 \pm \frac{c}{\sqrt{x_1}},
\end{equation}
where $c >0$ is an integration constant fixed as $1$ in the figure. The blue trajectory corresponds to the positive-$H$ blue curve of the left panel, a spacetime that begins in a singularity and evolves to de Sitter. The negative-$\dot{a}$ branch of the red trajectory corresponds to the negative-$H$ blue curve of the left panel, representing a spacetime that begins as Minkowski and collapses to a singularity. Finally, the positive-$\dot{a}$ branch of the red trajectory corresponds to the unphysical red curve of the left panel.

The instability of Minkowski space-time is a peculiarity of the linear dependence between $\rho_V$ and $H$. In order to see that, let us consider a general case with $\rho_V = 3 h(H)$, where $h(H)$ is an arbitrary function. As before, to avoid negative matter densities we will take $\dot{H} \leq 0$. We expect that vacuum decays into pressureless matter, and not the opposite, hence we have $h'(H) \geq 0$ (see Eq. (\ref{conservation})). In order to have zero vacuum density in flat space-time, we will consider functions such that $h(0) = 0$. As a further condition, we impose that $h'(H)$ does not diverge at $H=0$. The evolution equation for $H$ has now the general form
\begin{equation}
\dot{H} = g(H)  = \frac{3}{2} \left[ h(H) - H^2 \right].
\end{equation}
The critical points, $\dot{H} = 0$, are given by the roots of the algebraic equation $h(H) = H^2$, which includes the Minkowski solution $H = 0$. The  stability of this solution depends on the sign of
\begin{equation}
g'(H) = \frac{3}{2} \left[ h'(H) - 2 H \right]
\end{equation}
at $H = 0$. Since $h'(0) \geq 0$, we have $g'(0) \geq 0$. The Minkowski solution is unstable for $g'(0) > 0$, and this only occurs when $h(H) \propto H$ near the critical point, i.e. the linear case analised above.  Otherwise we have $g'(0) = 0$, and there is indifferent stability.

\section{Perturbations}

\begin{figure*}
\centerline{\includegraphics[height=4.9cm]{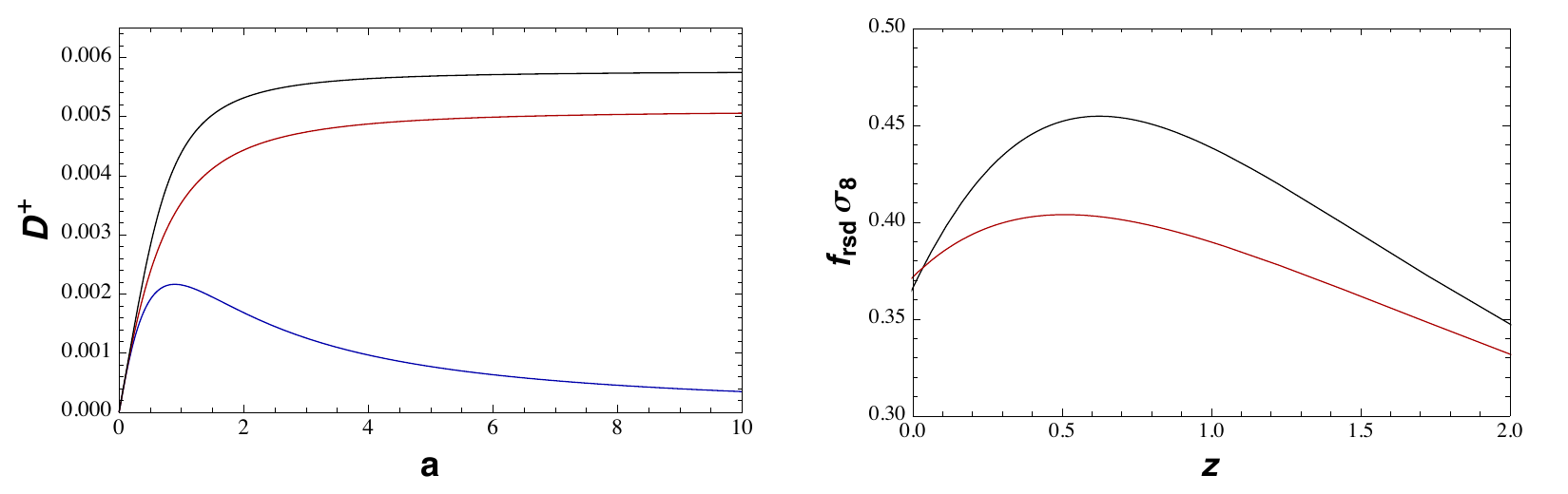} }
\caption{{\bf Left panel}: Evolution of the density contrast with the scale factor for the $\Lambda$CDM model (black curve) and for the interacting model: for dark matter (blue) and for baryons (red).
 {\bf Right panel}: Evolution of the baryonic $f_{rsd}\sigma_{8}$ versus $z$ for the $\Lambda$CDM model (black) and for the interacting model (red).}
\label{figb}
\end{figure*}

The reader has noticed the absence of baryons in the analyses above, where for simplicity we have considered a cosmic substratum formed by only interacting dark energy and dark matter. This is also a good approximation for testing the interacting model against background observations and the present matter power spectrum \cite{linearmodel,hermano}. However, the more realistic case in which conserved baryons are explicitly included leads to significant differences when the matter growing function is analysed. In Ref. \cite{Borges&Wands}, for example, the authors have pointed out a suppression in the dark matter growth rate owing to a corresponding late-time suppression in the dark matter density contrast. This suppression, caused by the homogeneous creation of dark matter from vacuum, also affects the total matter growth rate, since baryons are subdominant at late times. A signature of this suppression could in principle be found in redshift space distortions, expressed by the bias-independent quantity \cite{Borges&Wands}
\begin{equation}\label{frsd}
f_{rsd} \sigma_8 (z) = (f + g) \sigma_8(0) D^N_+,
\end{equation}
where $\sigma_8$ measures the amplitude of density fluctuations at $8$ Mpc, $D^N_+$ is the density contrast normalised to $1$ at $z = 0$, $f = D'_+/(\cal{H}D_+)$ (the prime means derivative w.r.t. conformal time, and ${\cal H} = Ha$) and $g = \Gamma/H$. The above equation comes from the relation between the matter peculiar velocity and its growing function \cite{Borges&Wands},
\begin{equation}\label{mar}
-\vartheta=(f+g) \, {\cal H}D_{+}.
\end{equation}
Needless to say, however, all we directly observe is baryonic matter. As baryons are conserved, there is not such a suppression in their density contrast. For conserved baryons $g = 0$ and we have, instead of ($\ref{mar}$), the standard relation between their peculiar velocity and growing function,
\begin{equation}\label{marb}
-\vartheta_b=f_b {\cal H} D^b_{+}.
\end{equation}
Hence, for baryonic matter we have, instead of (\ref{frsd}),
\begin{equation}\label{frsdb}
f_{rsd}(z)\sigma_{8}(z)=f_b\sigma_{8}(0)D^{N}_{b+}.
\end{equation}

This does not mean that the baryonic velocity field does not coincide with the dark matter one. Indeed, if we assume that vacuum is not perturbed, there is no momentum transfer to dark matter, and then it follows geodesics like baryons \cite{linearmodel}. This can be verified by taking the perturbed Euler equation for a pressureless interacting component $i$, which expresses its momentum conservation. In the Newtonian gauge it is written as
\begin{equation}
(\rho_i \vartheta_i)' + 4 {\cal H} \rho_i \vartheta_i +\rho_i \Phi = a Q_i \vartheta_i,
\end{equation}
where $Q_i$ is the interaction term, and $\Phi$ is the gravitational potential. Making use of the background conservation equation
\begin{equation} \label{background_conservation}
\rho'_i + 3{\cal H} \rho_i = -a Q_i,
\end{equation}
we can obtain, for both baryonic and dark matter components,
\begin{equation} \label{velocidades}
\vartheta_i' + {\cal H} \vartheta_i = - \Phi.
\end{equation}
Therefore, given the same initial conditions, dark matter and baryons follow the same geodesics. Note also that, as (\ref{velocidades}) is valid for any $Q_i$, it is also valid for total matter.

Relations (\ref{mar}) and (\ref{marb}) can also be derived in the Newtonian gauge, in the limit of sub-horizon scales $k \gg {\cal H}$. In this gauge, the energy conservation equation for a pressureless interacting component is given by
\begin{equation} \label{Newtonian_conservation}
\delta \rho'_i + 3{\cal H} \delta \rho_i - 3 \rho_i \Phi' + \rho_i \vartheta_i = -a Q_i \Phi.
\end{equation}
Using the background conservation equation (\ref{background_conservation}), we have
\begin{equation} \label{Newtonian}
 \delta'_{i} - \frac{aQ_i}{\rho_{i}} \delta_{i} + \vartheta_{i} = - \frac{aQ_i}{\rho_{i}} \Phi + 3 \Phi'.
\end{equation}
For the gravitational potential one can obtain, from the perturbed Einstein equations,
\begin{equation}
 k^2 \Phi =  -\frac{a^2}{2} \delta \rho_m - \frac{3 {\cal H} a^2 \rho_m}{2 k^2}\vartheta_m,
\end{equation}
where $\vartheta_m$ and $\rho_m$ are, respectively, the velocity potential and density of total matter. Taking the limit $k \gg {\cal H}$, we have the Poisson's equation
\begin{equation} \label{Poisson}
k^2 \Phi = -\frac{a^2}{2} \delta \rho_m \quad \quad (k \gg {\cal H}).
\end{equation}
Substituting into (\ref{Newtonian}) and taking again the small-scale limit, we obtain
\begin{equation}\label{con_i}
\delta_{i}'-\frac{aQ_i}{\rho_{i}}\delta_{i}+\vartheta_i=0 \quad \quad (k \gg {\cal H}).
\end{equation}
If we now take $Q_i/\rho_i = \Gamma$ for dark matter, $Q_i = 0$ for baryons and use the definitions above for $g$ and $f$, we obtain equations (\ref{mar}) and (\ref{marb}), respectively. On the other hand, by taking (\ref{background_conservation}) and (\ref{Newtonian_conservation}) for the total matter, it is easy to derive, in the same way,
\begin{equation}\label{mar_m}
-\vartheta_m =(f_ m+\tilde{g}){\cal H}D^{m}_+,
\end{equation}
where we defined $\tilde{g} \equiv (\rho_{dm}/\rho_m) g$.
In this way, besides to show the equality between the components velocities, we are proving the relations
\begin{equation}
f_b D^b_+ = (f_{dm} + g) D_+^{dm} = (f_m + \tilde{g}) D_+^m.
\end{equation}
The question now is how to choose between (\ref{frsd}) and (\ref{frsdb}) (or the total matter corresponding quantity) for contrasting observations. The choice will depend on how we normalise the density contrast today. As we have discussed above, the dark matter and total matter linear spectra suffer a late-time suppression that does not affect the baryonic spectrum. When we fix $\sigma_8(0) = 0.83$, we are taking the amplitude of the observed, baryonic spectrum, and (\ref{frsdb}) should be used\footnote{One may argue that $\sigma_8$ can also be inferred from weak lensing or CMB observations, which are sensitive to both baryonic and dark matter distributions. Actually, all these observations are directly sensitive to the gravitational potential generated by the total matter. It is possible to shown \cite{hermano} that, despite the suppression in the dark matter contrast, the gravitational potential is almost unaffected, because it is proportional to $\delta \rho_m$, not to $\delta_m$.}.

The background and perturbation solutions when conserved baryons are explicitly considered were studied in Ref. \cite{hermano}. The background is determined by the conservation and Friedmann equations\footnote{In equations (\ref{hermano})-(\ref{65}), the prime means derivative w.r.t. the scale factor $a$.}$^,$\footnote{We are assuming that total matter is created at a constant rate, which, in the presence of conserved baryons, is not the same as assuming that dark matter is created at a constant rate. The corresponding difference in the background solution is negligible, since baryons are subdominant.}
\begin{eqnarray} \label{hermano}
a H \rho'_m + 3H\rho_m &=& \Gamma \rho_m,\\
2aHH' + \rho_m &=& 0.
\end{eqnarray}
The perturbations equations are given, in the comoving-synchronous gauge, by
\begin{eqnarray} \label{64}
a^2 H^2 \delta''_m + aH (3H + aH' + \Gamma) \delta'_m &+& 2\Gamma H \delta_m  \nonumber \\ &=& \frac{1}{2} \rho_m \delta_m, \\ \label{65}
a^2 H^2 \delta''_b + aH (3H + aH') \delta'_b &=& \frac{1}{2} \rho_m \delta_m.
\end{eqnarray}
For sub-horizon scales they coincide with the Newtonian gauge equations. The reader may note the absence of creation terms (those proportional to $\Gamma$ in (\ref{64})) in the last equation, which determines the baryonic contrast. This does not mean that baryons are not affected by the creation of dark matter, since the latter is present in the source term on the right-hand side of (\ref{65}). However, the effect is indirect and the conserved baryons will not suffer the same suppression that affects dark matter.

In the left panel of Fig. 2 we show the evolution of the density contrasts of dark matter and baryonic matter, as well as the matter contrast in the $\Lambda$CDM model, for the same initial conditions. We can see that the baryon density contrast in the interacting model presents a small difference in its normalisation as compared to the standard one. Nevertheless, it does not suffer the strong late-time suppression observed in the dark matter contrast. While the latter goes to zero in the asymptotic future, the former tends to a constant like in the standard model. The resulting combination $f\sigma_8$ for baryons is shown in the right panel of Fig. 2, together with the $\Lambda$CDM one. In this figure we have used for the interacting model its best fit matter density parameter $\Omega_{m0} = 0.45$ \cite{linearmodel}. The two models give the same prediction for $z=0$, while in the past the relative difference between their curves is less than $12\%$. Let us remark that baryons are not coupled to radiation in the present analysis. Their coupling may, in principle, delay the contrast suppression for both dark matter and baryonic components.

\section{From \MakeLowercase{de} Sitter to \MakeLowercase{de} Sitter}

Let us now explore the limit of very high energies, with the vacuum density scaling as $\rho_V = 3 H^4$, where we have conveniently chosen the proportionality constant. Now, the Friedmann equations, given by
\begin{eqnarray} \label{Friedmann2}
3H^2 = \rho_R + \rho_V,\\ \label{conservation2}
\dot{\rho}_R + 4 H \rho_R = -\dot{\rho}_V
\end{eqnarray}
($\rho_R$ refers to relativistic matter), lead to the evolution equation
\begin{equation}\label{2nd}
\dot{H} = g(H) = 2H^4 - 2H^2.
\end{equation}
We have again two critical points, $H = 0$ and $H = 1$, but, since we are dealing with a high energy spacetime, only the latter has physical sense. We have $g'(1) = 4 > 0$, showing that $H = 1$ is an unstable de Sitter critical point. 

\begin{figure}
\includegraphics[height=3.5cm]{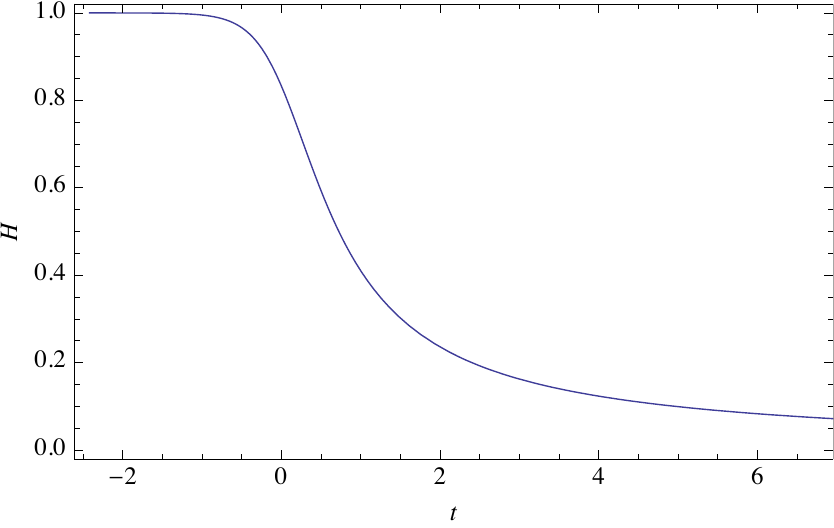}
\caption{$H$ as a function of $t$ in the high energy regime.}
\end{figure}

If we are restricted to a decreasing $H$ (in order to respect the condition $\rho_R > 0$), the solution of (\ref{2nd}), with the origin of time arbitrarily fixed, is given by
\begin{equation}
2t = \frac{1}{H} - \tanh^{-1}H.
\end{equation}
It is depicted in Fig. 2, and represents a universe evolving from a primeval de Sitter phase to a radiation-dominated one. As shown in \cite{Reza}, this transition, although remembering inflation in some aspects, does not generate a scale-invariant spectrum of primordial fluctuations, and a subsequent proper inflation period is still needed. Apart the standard inflation models based on self-interacting scalar fields, this can also be achieved if we assume the existence of an intermediate-scale regime in which the vacuum density decays as $\rho_V \propto H^2$, with a corresponding relativistic-matter creation rate $\Gamma \propto H$ \cite{Chimento}.

\label{highz}

\section{Conclusion}

Is there a cosmological constant problem? We commonly state that the vacuum density, regularised by a Planck scale cutoff, is huge as compared to the observed cosmological term. Even adopting lower cutoffs, as the energy scale of the QCD chiral condensation (the latest cosmological phase transition), the obtained order of magnitude is still too large. However, it is a matter of fact that in a flat background, where such a calculation is usually performed, nothing gravitates and the vacuum term should be trivially null. A less naive estimation can be tentatively made in a curved spacetime, with a normalisation procedure that correctly treats the flat-spacetime divergent contribution. In the case of a conformal scalar field in a de Sitter (or approximately de Sitter) background, this leads to a vacuum density of the order of $H^4$ \cite{conformal}. For our present Universe, this would lead to a tiny cosmological term, and the problem would be rather to explain why the observed cosmological constant is not zero. On the other hand, in the limit of very high energies this result implies an unstable de Sitter universe, as shown above. In this way, it provides a non-singular scenario in which a past-eternal de Sitter spacetime suffers a fast transition to a radiation dominated phase.

Speculations about the Universe's origin apart, the late-time normalised vacuum density can also be, in principle, evaluated and compared to observations. In the low energy limit, the contribution of interacting fields cannot be neglected, and estimations of the QCD vacuum has suggested a term scaling as $\Lambda \sim m^3 H$, where $m$ is the energy scale of the chiral phase transition \cite{Schutzhold}. It is noteworthy that this gives the observed figure of the cosmological term. Furthermore, a model with vacuum decaying linearly with $H$, producing dark matter at a constant rate, has shown a good concordance when tested against observations \cite{linearmodel,hermano}. In this short contribution we have shown that such a prescription for the vacuum decay also presents interesting features from a dynamical system viewpoint. A stable de Sitter universe is the future limit of any expanding solution. On the other hand, the trivial Minkowski solution is unstable and collapses to a singularity. Needless to say, those QFT estimations in curved spacetimes are something to be further developed and fully verified. Nevertheless, it would not be surprising if the Universe state and fate were ultimately determined by standard fundamental physics.


\section*{Acknowledgements}

We are thankful to J. S. Alcaniz for suggesting the problem and to C. Chirenti for helpful comments. SC is partially supported by CNPq.

\end{document}